\documentclass[conference]{IEEEtran}

\IEEEoverridecommandlockouts

\usepackage[utf8x]{inputenc}
\usepackage[T1]{fontenc}
\usepackage[american]{babel}
\usepackage[center]{caption}
\usepackage{graphicx}
\usepackage{listings}
\usepackage{float}
\usepackage{amsmath,amssymb,exscale}
\usepackage{blindtext, graphicx}
\usepackage{verbatim}
\usepackage{algorithm}
\usepackage{algpseudocode}
\usepackage{fancyvrb}
\usepackage{bera}
\usepackage{amsmath}
\usepackage{mathtools}

\usepackage{scalerel}
\usepackage{tikz}
\usetikzlibrary{svg.path}
\definecolor{orcidlogocol}{HTML}{A6CE39}
\tikzset{
  orcidlogo/.pic={
    \fill[orcidlogocol] svg{M256,128c0,70.7-57.3,128-128,128C57.3,256,0,198.7,0,128C0,57.3,57.3,0,128,0C198.7,0,256,57.3,256,128z};
    \fill[white] svg{M86.3,186.2H70.9V79.1h15.4v48.4V186.2z}
                 svg{M108.9,79.1h41.6c39.6,0,57,28.3,57,53.6c0,27.5-21.5,53.6-56.8,53.6h-41.8V79.1z M124.3,172.4h24.5c34.9,0,42.9-26.5,42.9-39.7c0-21.5-13.7-39.7-43.7-39.7h-23.7V172.4z}
                 svg{M88.7,56.8c0,5.5-4.5,10.1-10.1,10.1c-5.6,0-10.1-4.6-10.1-10.1c0-5.6,4.5-10.1,10.1-10.1C84.2,46.7,88.7,51.3,88.7,56.8z};
  }
}

\newcommand\orcid[1]{\href{https://orcid.org/#1}{\mbox{\scalerel*{
\begin{tikzpicture}[yscale=-1,transform shape]
\pic{orcidlogo};
\end{tikzpicture}
}{|}}}}

\usepackage[bookmarks=false]{hyperref}

\def\BibTeX{{\rm B\kern-.05em{\sc i\kern-.025em b}\kern-.08em
    T\kern-.1667em\lower.7ex\hbox{E}\kern-.125emX}}

\makeatletter 
\let\old@ps@headings\ps@headings 
\let\old@ps@IEEEtitlepagestyle\ps@IEEEtitlepagestyle 
\def\confheader#1{%
\def\ps@headings{%
\old@ps@headings%
\def\@oddhead{\strut\hfill#1\hfill\strut}%
\def\@evenhead{\strut\hfill#1\hfill\strut}%
}%
\def\ps@IEEEtitlepagestyle{%
\old@ps@IEEEtitlepagestyle%
\def\@oddhead{\strut\hfill#1\hfill\strut}%
\def\@evenhead{\strut\hfill#1\hfill\strut}%
}%
\ps@headings%
} 
\makeatother 

\usepackage{booktabs}
\usepackage{tabularx}
\usepackage{multirow}
\usepackage{stmaryrd}
\usepackage{enumitem}
\usepackage{xspace}

\floatstyle{boxed}
\newfloat{proto}{!ht}{lop}
\floatname{proto}{Protocol}

\newcommand{\sbline}{\\[.1\normalbaselineskip]\small}
\newcommand{\protosect}[1]{\small\textbf{#1}}

\newenvironment{protocol}[1]{
  \proto
    \tabularx{\linewidth}{ X }
      \textbf{Protocol: {\large #1}} \\
      \hline
      \small
}{\small\endtabularx\endproto}

\newcommand{\asynbit}{\ensuremath{\iota}\xspace}

\newcommand{\Zn}{\ensuremath{\mathbb{Z}_{N}}\xspace}
\newcommand{\Znn}{\ensuremath{\mathbb{Z}_{N^2}}\xspace}

\newcommand{\Zq}{\mathbb{Z}_q}

\newcommand{\BO}{\mathcal{O}}
\newcommand{\shareq}[1]{\ensuremath{\llbracket{#1}\rrbracket_{_q}}\xspace}
\newcommand{\hmu}[1]{\ensuremath{\hat{\mu}_{#1}}\xspace}
\newcommand{\hvar}[1]{\ensuremath{\hat{Var}(#1)}\xspace}

\newcommand{\hv}[1]{\ensuremath{\hat{\theta_\mathsf{#1}}}\xspace}
\newcommand{\lv}[2]{\ensuremath{#1_\mathsf{#2}}\xspace}

\newcommand{\mbf}[1]{\mathbf{#1}}
\newcommand{\pmpc}{\ensuremath{\pi_{\hat{\mu}^\mathsf{MPC}}}\xspace}
\newcommand{\pmhe}{\ensuremath{\pi_{\hat{\mu}^\mathsf{HE}}}\xspace}

\newcommand{\padd}{\ensuremath{\pi_{\mathsf{ADD}}}\xspace}
\newcommand{\pmul}{\ensuremath{\pi_{\mathsf{MUL}}}\xspace}

\newcommand{\party}[1]{\ensuremath{\mathcal{P}_{{#1}}}\xspace}
\newcommand{\parti}{\party{i}}

\usepackage{array}
\newcommand{\PreserveBackslash}[1]{\let\temp=\\#1\let\\=\temp}
\newcolumntype{C}[1]{>{\PreserveBackslash\centering}p{#1}}
\newcolumntype{R}[1]{>{\PreserveBackslash\raggedleft}p{#1}}
\newcolumntype{L}[1]{>{\PreserveBackslash\raggedright}p{#1}}

\begin{document}

\title{Monte Carlo execution time estimation for Privacy-preserving Distributed Function Evaluation
protocols}

\author{
\IEEEauthorblockN{
    Stefano M P C Souza\IEEEauthorrefmark{1} \orcid{0000-0002-7935-2580},
    Daniel G Silva \orcid{0000-0003-2858-1110}
}
\IEEEauthorblockA{
    Department of Electrical Engineering, University of Bras\'ilia (UnB), Bras\'ilia, Brazil
\\
    Email: stefanomozart@ieee.org\IEEEauthorrefmark{1}}
}

\IEEEoverridecommandlockouts 
\IEEEpubid{\makebox[\columnwidth]{Copyright Notice } 
\hspace{\columnsep}\makebox[\columnwidth]{ }} 

\maketitle

\begin{abstract}

Recent developments in Machine Learning and Deep Learning depend heavily on cloud computing and
specialized hardware, such as GPUs and TPUs. This forces those using those models to trust private
data to cloud servers. Such scenario has prompted a large interest on Homomorphic Cryptography and
Secure Multi-party Computation protocols that allow the use of cloud
computing power in a privacy-preserving manner.

When comparing the efficiency of such protocols, most works in literature resort to complexity
analysis that gives asymptotic higher-bounding limits of computational cost when input size tends to
infinite. These limits may be very different from the actual cost or execution time, when performing
such computations over small, or average-sized datasets.

We argue that Monte Carlo methods can render better computational cost and time estimates, fostering
better design and implementation decisions for complex systems, such as Privacy-Preserving Machine
Learning Frameworks.

\end{abstract}


\section{Introduction}

\IEEEPARstart{M}{achine learning} (ML) applications are revolutionizing industry, education, and
many other fields. Recent developments of ML, especially Deep Learning (DL) methods and algorithms,
have shown remarkable results in a wide range of applications: from computational vision to text
generation \cite{Wilson:GANN}. 

As new ML algorithms increase in complexity and computing cost, developments in the area extend 
their dependency on cloud computing and hardware accelerated functions found on specialized
chipsets, such as Graphical Processing Unities (GPUs) and Google's Tensor Processing Units (TPUs) \cite{LeCun}.

This move towards the cloud forces individuals and organizations to trust their private data to
external servers in order to harness the predictive power of such algorithms. Electronic Health
Records and tax information are examples of sensitive data that are extremely relevant for
predictive modeling but are also legally bound to strong privacy-preserving guarantees. Many 
information-theoretical and cryptographic protocols were developed in order to allow secure and
privacy-preserving delegation of computation. Most are based on homomorphic encryption systems (HE) 
or secret sharing Secure Multi-Party Computation (MPC) protocols.

In order to compare the efficiency of these protocols, researchers usually resort to theoretical
complexity analysis, that provides for asymptotic higher-bounding limits of computational cost when
the input size tends to infinite \cite{Naor}. These limits, drafted from the the underlying
mathematical primitives, often drive design and implementation decisions on large and complex
systems. Ultimately, they may determine important investment in research and development of such
protocols. Nonetheless, theoretical limits may be very different from the actual average 
computational cost and execution times observed when executing the protocols over small, or
medium-sized datasets.

Probabilistic and ML models are commonly used to estimate cost, execution time and other statistics
over complex systems and algorithms, especially in control or real-time systems engineering
\cite{daniel, Iqbal}.

We propose the use of Monte Carlo methods in order to estimate execution times for 
privacy-preserving computations, considering different protocols and input sizes. Section II 
presents a short review on privacy-preserving computation and its cost. Section III has a brief
description of the Monte Carlo methods used. Section IV discusses implementation details and results
of the various Monte Carlo experiments we performed. Finally, Section V presents key conclusions and
points out relevant questions open for further investigation.

\section{Privacy-Preserving Computation}\label{sec:02}
As privacy-preserving computation grew in importance and attention, many Privacy-Preserving Machine
Learning (PPML) and Privacy-Preserving Function Evaluation (PPFE) frameworks have been developed.
The protocols that form the basic building blocks of these frameworks, are usually based on 
homomorphic cryptography primitives and Secure Multi-Party Computation protocols. Some of the first
frameworks to appear in literature, for instance, used MPC protocols based on Secret Sharing. Among
those preceding results are FairPlayMP \cite{Ben-David:FairPlay} and Sharemind 
\cite{Bogdanov:Sharemind}.

Recent developments include frameworks like PySyft \cite{Ryffel:PySyft}, that uses HE protocols, and
Chameleon \cite{Riazi:Chameleon}, that uses MPC for linear operations and Garbled Circuits (a form
of HE) for non-linear evaluations. Other MPC frameworks in literature include PICCO 
\cite{Zhang:PICCO}, TinyGarble \cite{Songhori:TinyGarble}, $\text{ABY}^3$ \cite{Demmler:ABY3} and
SecureML \cite{Mohassel:SecureML}.

\subsection{Homomorphic Cryptography}\label{sec:02:he}
A cryptosystem is said to be homomorphic if there is an homomorphism between the domain (the message
space $\mathcal{M}$) and the image (the cipher space $\mathcal{C}$) of its encryption function 
$Enc(m)$ \cite{eu:CF}. An homomorphism is a map from one algebraic structure to another, that
maintains its internal properties. So, if there is an internally well defined relation, or function,
in $\mathcal{M}$, $f_\mathcal{M}: \mathcal{M} \rightarrow \mathcal{M}$, there will be a 
corresponding function defined in $\mathcal{C}$, $f_\mathcal{C}: \mathcal{C} \rightarrow
\mathcal{C}$, such that:
\begin{gather*}
    \forall m \in \mathcal{M}, \\
    f_{\mathcal{C}}(Enc(m)) \equiv Enc(f_{\mathcal{M}}(m))
\end{gather*}

Fully Homomorphic Encryption (FHE) refers to a class of cryptosystems for which the homomorphism is
valid for every function defined in $\mathcal{M}$. That is:
\begin{gather*}
    \forall f_{\mathcal{M}}:\mathcal{M}\rightarrow\mathcal{M}, \\
    \exists f_{\mathcal{C}}:\mathcal{C}\rightarrow\mathcal{C} ~| ~f_{\mathcal{C}}(Enc(m)) \equiv
    Enc(f_{\mathcal{M}}(m)) \hspace{10pt}
\end{gather*}

The most commonly used homomorphic cryptography systems, however, are only partially homomorphic.
There are additive homomorphic systems, multiplicative homomorphic systems and systems that combine
a few homomorphic features. For example, Paillier's cryptosystem has additive and multiplicative
homomorphic features that can be used to delegate a limited set of computations over a dataset,
without compromising its confidentiality \cite{Paillier, eu:CPE}.

The underlying primitive in Paillier's system is the quadratic residuosity problem. Its construction
harnesses the homomorphism between the fields \Zn and \Znn to render the following features:
\begin{description}[leftmargin=1em]
  \item[Asymmetric cryptography:] it is possible to perform homomorphic computations over the 
  encrypted data using the public key. Knowing the results of the computation, nevertheless, requires
  access to the private key;
  \item[Additive homomorphism:] the multiplication of two ciphertexts equals the ciphertext of 
  the sum of their respective messages. That is:
  \begin{equation*}
    Enc(m_{1}).Enc(m_{2}) ~\text{mod~} N^{2} = Enc(m_{1} + m_{2} ~\text{mod~}\xspace N)
  \end{equation*}
  \item[Multiplicative homomorphism:] a ciphertext to the power of an integer equals the 
  ciphertext of the multiplication of the original message by that integer. That is:
  \begin{equation*}
    Enc(m_{1})^{m_{2}} ~\text{mod~} N^2= Enc(m_{1} . m_{2} ~\text{mod~} N)
   \end{equation*}
\end{description}

\subsection{Secret Sharing}
Introduced by Yao \cite{Yao:MPC}, Secure Multi-Party Computation refers to a set of protocols and
algorithms that allow a group of computing parties $\mathcal{P}$ to evaluate a function $F(X)$ over 
a set $X = (x_i, x_2, ..., x_n)$ of private inputs, in a way that guarantees participants gain
knowledge only on the global function result, but not on each others inputs.

\begin{protocol}{\padd}
  \protosect{Input:} Secret shares $\shareq{x_1}, \dots, \shareq{x_n}$
  \sbline 
  \protosect{Output:} $\shareq{z} = \sum_{i=1}^n \shareq{x_i}$
  \sbline 
  \protosect{Execution:}
  \begin{enumerate}
    \item Each party $\parti \in \mathcal{P}$ computes $z_i = x_i + y_i$
    \item Each party $\parti$ broadcasts $z_i$
  \end{enumerate}
  \caption{\small Secure Distributed Addition Protocol \padd}
  \label{prot:add}
\end{protocol}

Additive secret sharing is one way to implement MPC. Protocol parties have additive shares of their 
secret values and perform joint computations over those shares. For example, to create $n$ additive
shares of a secret value $x \in \mathbb{Z}_q$, a participant can draw $(x_1,\ldots,x_n)$ uniformly
from $\{0,\ldots,q-1\}$ such that $x = \sum_{i=1}^{n}x_i \mod q$. We denote this set of shares by 
$\shareq{x}$.

Notice that access to any proper subset of $\shareq{x}$ gives no information about $x$. A shared
secret can only be revealed after gathering all shares. Likewise, the result of a protocol executing
linear transformations over such shares can only be known if all the local results at each computing
party are combined.

Given two sets of shares $\shareq{x}$, $\shareq{y}$ and a constant $\alpha$, it is trivial to 
implement a protocol like \padd or \pmul in order to locally compute linear functions over the sum
of their respective shares and broadcast local results to securely compute values such as
$\shareq{z} = \alpha (\shareq{x} \pm \shareq{y})$, $\shareq{z} = \alpha \pm (\shareq{x} \pm
\shareq{y})$, and $\shareq{z} = \alpha_1 (\alpha_2 \pm (\shareq{x} \pm \shareq{y}))$. In all those
operations, $\shareq{z}$ is the shared secret result of the protocol. The real value of $z$ can 
only be obtained if one has knowledge of the local $z_i$ values held by all computing parties, 
and performs a last step of computation to obtain $z = \sum_{i=1}^{n} z_i \mod q$.

\begin{protocol}{\pmul}
  \protosect{Setup:} 
  \begin{enumerate}
      \item The Trusted Initializer draws $\mbf{u}, \mbf{v}, \mbf{w}$ uniformly from $\Zq$, such 
      that $\mbf{w}=\mbf{uv}$ and distributes shares $\shareq{\mbf{u}}, \shareq{\mbf{v}}$ and
      $\shareq{\mbf{w}}$ to protocol parties
      \item The TI draws $\asynbit \xleftarrow{u} \{1,\dots,n\}$, and sends asymmetric bit \{1\} to
      party \party{\asynbit} and \{0\} to parties \party{i\neq \asynbit}
  \end{enumerate}
  \sbline
  \protosect{Input:} Shares $\shareq{x}$, $\shareq{y}$
  \sbline
  \protosect{Output:} $\shareq{z} = \shareq{xy}$
  \sbline
  \protosect{Execution:}
  \begin{enumerate}
    \item Each party \parti locally computes $d_i \gets x_i - \mbf{u}_i$ and $e_i \gets y_i - 
    \mbf{v}_i$
    \item Parties broadcast $d_i, e_i$
    \item Each party computes $d \gets \sum_{i=1}^{n} d_i, e \gets \sum_{i=1}^{n} e_i$
    \item The party \party{\asynbit} holding the asymmetric bit computes $z_\asynbit \leftarrow
    \mbf{w}_\asynbit + d\mbf{v}_\asynbit + e\mbf{u}_\asynbit + de$
    \item All other parties \party{i\neq \asynbit} compute $z_i \gets \mbf{w}_i + d\mbf{v}_i +
    e\mbf{u}_i$
  \end{enumerate}
  \caption{Secure Distributed Multiplication \pmul}
  \label{prot:distmult}
\end{protocol}

\subsection{The cost of privacy-preserving computation}
The aforementioned PPML frameworks are proof of the research interest in such methods and the large
investment put in the development of privacy-preserving computation techniques. It is important
to note that some of the most recent and relevant results are sponsored by major cloud service
providers, especially those with specific Machine Learning related cloud services, such as Google,
IBM and Amazon.

Therefore, not only the capabilities of each protocol or framework are of great importance, but 
also their economical viability: determined mostly by their computational cost. Expected execution 
time and power consumption may drive decisions that have impact on millions of users and on relevant 
fields of application. In spite of the importance of the efficiency of their solutions, authors
usually only discuss very briefly the estimated computational cost or execution times. 

Very few works on PPML publish their results with computing times observed against benchmark 
datasets/tasks (e.g. classification on the ImageNet or the Iris dataset). When any general estimate
measure is present, it is usually the complexity order $\BO(g(n))$, which defines an asymptotically
lower-bounding asymptotic complexity or cost function $g(n)$. 

The order function, $g(n)$, represents the general behaviour or shape of a class of functions. So, 
if $t(n)$ is the function defining the computational cost of the given algorithm over its input size
$n$, then stating that $t(n) \in \BO(g(n))$ means that there is a constant $c$ and a large enough
value of $n$ after which $t(n)$ is always less than $c \times g(n)$.

\begin{equation*}
  t(n) \in \BO(g(n)) \longrightarrow \exists ~c, k ~| ~\forall ~n \geq k, t(n) \leq cg(n)  
\end{equation*}

A protocol is usually regarded efficient if its order function is at most polynomial. 
Sub-exponential or exponential orders, on the other hand, are usually deemed prohibitively high. 

For example, the authors of $\text{ABY}^3$ \cite{Demmler:ABY3} assert that their Linear Regression
protocol is the most efficient in literature, with cost $\mathcal{O}(B+D)$ per round, where $B$ is a
training batch size and $D$ is the feature matrix dimension \cite{Demmler:ABY3}. That may seem like 
a very well behaved linear function over $n$, which would lead us to conclude they devised an 
exceptionally efficient protocol with order $\BO(n)$.

Nevertheless, this order expression only gives a bound on the number of operations to be performed
by the algorithm. However, it does not inform an accurate estimate for execution time. And, more
importantly, the order function will only bound the actual cost function for extremely large input
sizes. Recall that $t(Kn) \in \BO(n)$, regardless of how arbitrarily large the constant $K$ may be.

Thus, for small input sizes, the actual cost may be many orders of magnitude higher than the
asymptotic bound. The addition protocol in \cite{Agarwal:Brain} is also of order $\BO(n)$, but one
would never assume that a protocol with many matrix multiplications and inversions can run as fast 
as the one with a few simple additions.

\section{Monte Carlo methods for integration}\label{sec:03}
There is another way to estimate execution times. All examples found in literature of 
privacy-preserving computation have one thing in common: their protocols depend heavily on
pseudo-randomly generated numbers, used to mask or encrypt private data. 

Those numbers are assumed to be drawn independently according to a specific probability density
function. That is, the algorithms use at least one random variable as input. Although the observed
execution time does not depend directly on the random inputs, it is directly affected by their
magnitude, or more specifically, their average bit-size.

Also, the size of the dataset has direct impact on execution time. If we consider the size of the
dataset as a random variable, then the interaction between the magnitude of the random numbers and
the numerical representation of the dataset are, unequivocally, random variables. So, it is safe
to assume that protocol runtimes, that are a function of the previous two variables, are also random
variables.

\begin{table*}[h!]
\caption{Experiment Type I - Runtimes in Milliseconds}\label{tab:exp1}
\scriptsize
\hfill{}
\begin{tabular}{l c l r r r r}
\toprule
Dataset & Protocol & M & \hv{cli} & \hvar{\hv{cli}} & \hv{srv} & \hvar{\hv{srv}} \\
\hline
\multirow{4}{8.5em}{{Dow Jones Index (750 instances)}} & \multirow{2}{3em}{\pmhe} 
{} & 1000 & 1943.94 & 4.36 & 13.03 & 0.001  \\
{} &{}  & 5000 & 1931.49 & 0.12 & 12.73 & 4.77$e^{-05}$\\
\cline{2-7}
{} & \multirow{2}{3em}{\pmpc} & 1000 & 134.35 & 6.66 & 0.25 & 3.15$e^{-05}$ \\
{} & {} & 5000 & 130.62 & 1.23 & 0.241 & 1.04$e^{-06}$ \\
\hline
\multirow{4}{8.5em}{Bank Marketing (4521 instances)} & \multirow{2}{3em}{\pmhe} 
{} & 1000 & 11630.08 & 165.05 & 24.75 & 0.006  \\
{} & {} & 5000 & 11514.43 & 24.22 & 24.44 & 5.05$e^{-04}$ \\
\cline{2-7}
{} & \multirow{2}{3em}{\pmpc} & 1000 & 289.51 & 6.96 & 0.89 & 1.63$e^{-04}$ \\
{} & {} & 5000 & 280.97 & 1.30 & 0.82 & 9.91$e^{-06}$ \\
\hline
\end{tabular}
\hfill{}
\end{table*}

Monte Carlo methods are a class of algorithms based on repeated random sampling that render
numerical approximations, or estimations, to a wide range of statistics, as well as the associated
standard error for the empirical average of any function of the parameters of interest. We know, for
example, that if $X$ is a random variable with density $f(x)$, then the mathematical expectation of
the random variable $T = t(X)$ is:
\begin{equation}
  E[t(X)] = \int_{-\infty}^{\infty} t(x)f(x)dx.
\end{equation}

And, if $t(X)$ is unknown, or the analytic solution for the integral is hard or impossible, then we
can use a Monte Carlo estimation for the expected value. It can can be obtained with:
\begin{equation}
  \hat{\theta} = \frac{1}{M} \sum_{i=1}^M t(x_i)f(x_i)
\end{equation}

In other words, if the probability density function $f(x)$ has support on a set $\mathcal{X}$, (that
is, $f(x) \geq 0 ~\forall x \in \mathcal{X}$ and $\int_\mathcal{X} f(x) = 1$), we can estimate the
integral 

\begin{equation}
  \theta = \int_\mathcal{X} t(x)f(x)dx
\end{equation}

by sampling $M$ instances from $f(x)$ and computing

\begin{equation}
  \hat{\theta} = \frac{1}{M} \sum_{i=1}^M t(x_i)
\end{equation}

The intuition is that, as $M$ grows, $X = \{x_1, \dots, x_m\}$ sampled from $\mathcal{X}$, the
support interval of $f(x)$, becomes closer to $\mathcal{X}$ itself. Therefore, the estimation
$\hat{\theta}$ will converge to the expected value $\theta$. This comes from the fact that the sample
mean is an unbiased estimator for the expected value.

We know that, by the Law of Large Numbers, the sample mean $\hat{\theta}$ converges to 
$E[\hat{\theta}] = \theta$ as $M \rightarrow \infty$. Therefore, we are assured that for a
sufficiently large $M$, the error $\varepsilon = E[\hat{\theta}]-\hat{\theta}$ becomes negligible. 

The associated variance for $\hat{\theta}$ is $Var(\hat{\theta}) = \sigma^2/M$, where $\sigma^2$ is
the variance of $t(x)$. Since we may not know the exact form of $t(x)$, or the corresponding mean and
variance, we can use the following approximation:
\begin{equation}
  \hat{Var}(\hat{\theta}) = \frac{1}{M^2}\sum_{i=1}^{M} \left(t(x_i) -\hat{\theta} \right)^2
\end{equation}

In order to improve the accuracy of our estimation, we can always increase $M$, the divisor in the
variance expression. That comes, however, with increased computational cost. We explore this 
trade-off in our experiments by performing the simulations with different values of $M$ and then
examining the impact of $M$ on the observed sample variance and on the execution time of the
experiment.

\section{Simulating a simple ``Mean Protocol''}\label{sec:04}
In order to simulate the scenario where the data owner wants to delegate the computation of some 
statistic over a private dataset, we implemented two simple protocols that compute the floor of the
mean of a list of numbers. Let $X = {x_0, \dots, x_\ell}$ be the input of the protocols, then the
output is $\hat{\mu} = \lfloor \frac{1}{\ell} \sum_{i=1}^nx_i \rfloor$. 

Both protocols have a client delegating a distributed computation to a fixed number of servers. The
\pmhe protocol uses homomorphic encryption. The client encrypts the instances in the dataset and
sends the ciphertexts to the servers. The servers perform a series of homomorphic additions and
one final homomorphic multiplication to multiply the sum by $\ell^{-1} \mod N$ (the modular
multiplicative inverse of the size of the dataset).

The \pmpc protocol is a composition of the Distributed Addition Protocol \padd, to sum the instances,
with the Distributed Multiplication Protocol \pmul, used to multiply that sum to $\ell^{-1}$. The
server part is performed by a fixed number $k$ of computing parties. The client generates $k$ shares
for each instance on the dataset and sends a set of $\ell$ shares to each computing party. The parties
will respond with \shareq{\mu}. The client then computes $\hmu = \frac{1}{k} \sum_{i=1}^k\mu_i \mod
q$. In our implementation, the client also performs the protocol steps designed to be executed by the
Trusted Initializer (as expressed in \padd and \pmul).

\begin{table*}[ht]
\begin{center}
\caption{Experiment II - Runtime in Milliseconds}\label{tab:exp2}
\scriptsize
\begin{tabular}{C{1.1cm} C{0.8cm} C{2em} | R{2.7em} R{3em} R{2.7em} R{3.9em} | R{2.7em} R{3em} R{2.7em} R{3.9em} | R{2.7em} R{3em} R{2.7em} R{3.7em} } 
\toprule
\multirow{3}{*}{Distribution} & {} & {} & \multicolumn{12}{c}{Dataset Size} \\ 
{} & {} & {} & \multicolumn{4}{c|}{50} & \multicolumn{4}{c|}{500} & \multicolumn{4}{c}{1000} \\
{} & Protocol & M & \hv{cli} & \hvar{\hv{cli}} & \hv{srv} & \hvar{\hv{srv}} & \hv{cli} & \hvar{\hv{cli}} & \hv{srv} & \hvar{\hv{srv}} & \hv{cli} & \hvar{\hv{cli}} & \hv{srv} & \hvar{\hv{srv}} \\
\midrule
\multirow{4}{*}{\shortstack{Uniform\\(80, 160)}} & \multirow{2}{2em}{\pmhe} & 1000 & 
184.73 & 0.88 & 12.51 & 0.010 & 1272.79 & 0.34 & 11.66 & 1.41$e^{-4}$ & 2555.45 & 12.77 & 13.43 & 9.99$e^{-4}$  \\
{} & {} & 5000 & 
167.58 & 0.06 & 11.57 & 7.19$e^{-4}$ & 1249.13 & 0.89 & 11.50 & 1.57$e^{-5}$ & 2510.23 & 5.49 & 13.24 & 1.05$e^{-4}$ \\
\cline{2-15}
{} & \multirow{2}{2em}{\pmpc} & 1000 & 
116.11 & 8.73 & 0.090 & 1.61$e^{-4}$ & 122.28 & 7.39 & 0.195 & 5.65$e^{-6}$ & 142.88 & 6.63 & 0.273 & 1.09$e^{-5}$ \\
{} & {} & 5000 & 
109.36 & 1.44 & 0.062 & 1.27$e^{-4}$ & 118.69 & 1.29 & 0.171 & 6.25$e^{-7}$ & 139.51 & 1.33 & 0.256 & 2.47$e^{-6}$ \\
\hline
\multirow{4}{*}{\shortstack{Normal\\(120, 30)}} & \multirow{2}{2em}{\pmhe} & 1000 & 
182.10 & 0.79 & 12.53 & 0.009 & 1273.84 & 0.38 & 11.67 & 8.66$^{-6}$ & 2551.78 & 11.26 & 13.44 & 0.001 \\
{} & {} & 5000 & 
167.26 & 0.06 & 11.55 & 4.67$^{-4}$ & 1248.96 & 0.085 & 11.05 & 3.10$^{-5}$ & 2508.26 & 1.49 & 13.22 & 1.02$e^{-4}$ \\
\cline{2-15}
{} & \multirow{2}{*}{\pmpc} & 1000 & 
119.31 & 9.52 & 0.073 & 4.36$e^{-5}$ & 118.09 & 5.92 & 0.195 & 3.62$^{-6}$ & 142.40 & 6.68 & 0.279 & 1.28$e^{-5}$ \\
{} & {} & 5000 & 
108.58 & 1.47 & 0.065 & 4.12$e^{-6}$ & 118.65 & 1.22 & 0.169 & 5.56$e^{-7}$ & 139.28 & 1.28 & 0.253 & 1.75$e^{-6}$ \\
\hline
\multirow{4}{*}{\shortstack{Gamma\\(2, 2)}} & \multirow{2}{*}{\pmhe} & 1000 & 
182.83 & 0.802 & 12.46 &0.0123 & 1273.60 & 0.41 & 11.66 & 1.09$^{-4}$ & 2555.44 & 12.23 & 13.49 & 0.001 \\
{} & {} & 5000 & 
167.29 & 0.064 & 11.57 & 5.53$^{-4}$ & 1248.99 & 0.10 & 11.50 & 1.65$^{-5}$ & 2507.93 & 1.46 & 13.24 & 1.13$e^{-4}$ \\
\cline{2-15}
{} & \multirow{2}{*}{\pmpc} & 1000 & 
114.85 & 7.61 & 0.080 & 1.21$^{-4}$ & 126.53 & 7.22 & 0.197 & 4.04$e^{-6}$ & 142.82 & 6.89 & 0.277 & 1.29$e^{-5}$ \\
{} & {} & 5000 & 
109.37 & 1.43 & 0.066 & 7.63$^{-6}$ & 118.18 & 1.19 & 0.170 & 6.26$e^{-7}$ & 138.03 & 1.25 & 0.253 & 1.93$e^{-6}$ \\
\hline
\multirow{4}{*}{\shortstack{Beta\\(30,2)}} & \multirow{2}{*}{\pmhe} & 1000 & 
182.41 & 0.712 & 12.64 & 0.0117 & 1275.10 & 0.38 & 11.66 & 9.55$e^{-5}$ & 2557.53 & 13.57 & 13.42 & 0.001 \\
{} & {} & 5000 & 
167.24 & 0.058 & 11.55 & 5.06$^{-4}$ & 1249.87 & 0.09 & 11.50 & 2.09$e^{-5}$ & 2511.68 & 5.25 & 13.23 & 1.09$e^{-4}$ \\
\cline{2-15}
{} & \multirow{2}{*}{\pmpc} & 1000 & 
120.52 & 9.53 & 0.092 & 3.53$^{-4}$ & 121.92 & 6.74 & 0.194 & 3.51$e^{-6}$ & 147.29 & 7.75 & 0.277 & 1.28$e^{-5}$ \\
{} & {} & 5000 & 
110.24 & 1.49 & 0.064 & 4.61$^{-6}$ & 118.20 & 1.21 & 0.171 & 1.21$e^{-6}$ & 138.79 & 1.21 & 0.254 & 2.44$e^{-6}$ \\
\bottomrule
\end{tabular}
\end{center}
\end{table*}

\subsection{Experiment Type I}
Using \pmhe and \pmpc, we perform two types of experiment. The first type consists in running the
protocols for $M \in \{1000, 5000\}$ iterations with fixed datasets. We use the Dow Jones Index
dataset from \cite{dow}, with 750 instances, to compute the mean of the `balance' feature. We also
use the short Bank Marketing dataset from \cite{bank}, with 4521 instances, to compute the mean of
the `volume' feature.

We know that each \pmhe protocol run will sample at least $\ell$ (the size of the dataset) instances
from the uniform distribution $Uniform(1, n)$ (for the encryption under Paillier's system). We are
running the MPC protocols with 3 computing parties. Thus, each \pmpc run will sample at least $3\ell$
instances from $Uniform(0, q)$, as defined in \padd.

For each protocol run, we record the execution times for client and server: \lv{t}{cli}, \lv{t}{srv}.
We consider that these values are observations from $\lv{t}{cli}(h(U))$ and $\lv{t}{srv}(h(U))$,
functions that give the runtimes for client and server, respectively, given $U = \{u_i, \dots,
u_\ell\}$ - a sample from the uniform distributions used internally by the protocols. 

Recall that we do not regard runtime as a direct function of $U$, but rather of a measure of 
magnitude of $u \in U$. So let $h(U)$ be, for example, a function with image on $\mathbb{R}$, that
gives the average position of the most significant bit in the unsigned integer representation of each
$u_i \in U$. It is clear that the runtime function $t(h(U)))$ has its domain on a Real interval. For
readability, we will suppress the notation of function composition ($t(h(U))$) and write
$\lv{t}{cli}(u)$ and $\lv{t}{srv}(u)$ from now on.

Now, we want to estimate:
\begin{gather}
  \lv{\theta}{cli} = \int \lv{t}{cli}(u)f(u)du \\
  \lv{\theta}{srv} = \int \lv{t}{srv}(u)f(u)du
\end{gather}

So we use the following Monte Carlo approximations:
\begin{gather}
    \hv{cli} = \frac{1}{M} \sum_{i=1}^{M} \lv{t}{cli}^{(i)} \\
    \hv{srv} = \frac{1}{M} \sum_{i=1}^{M} \lv{t}{srv}^{(i)}
\end{gather}

We present the results in Table \ref{tab:exp1}, for $M \in \{1000, 5000\}$, along with the associated
variances.

\subsection{Experiment Type II}
In the second batch of experiments, we wanted to examine the influence of the probability 
distribution of the values in the dataset on runtimes. We modeled an experiment where datasets of
sizes $\ell \in \{50, 100, 1000\}$ are sampled from $U(80, 120)$, $N(120, 30)$ and the 
r.v.'s $G' = 120*G$, where $G \sim Gamma(2,2)$, and $B' = 120*B$, where $B \sim Beta(30, 2)$.

As we did in the previous series of experiments, we record \lv{t}{cli}, \lv{t}{srv}. This time, we
consider that these values are observations from $\lv{t}{cli}(X)$ and $\lv{t}{srv}(X)$, the
functions that render the runtimes for client and server, respectively, given the random dataset $X$.
So, let $f(x)$ be the density of $X$, we want to estimate 

\begin{gather}
    \lv{\theta}{cli} = \int \lv{t}{cli}(x) f(x)dx \\
    \lv{\theta}{srv} = \int \lv{t}{srv}(x) f(x)dx
\end{gather}

Notice that these have the same form of the integrals estimated in the first experiment. Hence, we 
use the same form for the Monte Carlo approximations:
\begin{gather}
    \hv{cli} = \frac{1}{M} \sum_{i=1}^{M} \lv{t}{cli}^{(i)} \\ 
    \hv{srv} = \frac{1}{M} \sum_{i=1}^{M} \lv{t}{srv}^{(i)}
\end{gather}

The estimations produced in the second group of experiments are listed on Table \ref{tab:exp2}, 
along with the associated variances.

\subsection{Implementation details}
Our experiments were written in Go. For the modular arithmetic, we used the native 
arbitrary-precision arithmetic library "math/big". To simulate the distributed computation and
communication, we used goroutines, a lightweight form of thread managed by the Go runtime, and Go
channels (typed sockets used for communication between goroutines).

All experiments were compiled with Go 1.12.3, on a Linux kernel 4.9.0-9-amd64, and ran on a 
Intel\textregistered\ Core\texttrademark\ i7-6500U CPU, with 4 cores at 2.50GHz clock. The code for
the experiments is open-sourced at the public git repository
\url{https://github.com/stefanomozart/montecarlo-protocol-runtime}. 

\subsection{Results}
Table  \ref{tab:exp1} shows the approximated values for \lv{t}{cli} and \lv{t}{srv} for \pmhe and 
\pmpc running with two datasets with different sizes and different magnitudes, as detailed in the 
previous section. It also brings the associated variance for those estimates for simulations with 
1000 and 5000 iterations.

Table \ref{tab:exp2} shows the values for \lv{t}{cli} and \lv{t}{srv}, and the respective variances, 
when the datasets vary in size and in density distribution.

The results of both experiments, type I and type II, confirm that increasing the number of simulation
iterations reduces significantly the variance of our estimator. In Table II, we can see that this reduction of variance is more noticeable for the larger datasets. We can affirm that,
for the given scenario, $M = 5000$ iterations produce an acceptable estimation.

The results in Table II also indicate that, for both protocols, the runtime may be a linear function 
of the size of the dataset. Fitting a simple model $\lv{t}{cli} \sim \beta_0 + \beta_1\ell$ (with 
R's implementation of the QR regression method), we get $\beta_0 = 18.479$, and $\beta_1 = 2.849$ for
the \pmhe protocol, and $\beta_0 = 107.317$, and $\beta_1 = 0.0414$ for \pmpc. The fitted $p$-values
are all smaller then $2.2e^{-16}$ for both models. This suggests that the runtime for the \pmhe
protocol is more sensitive to variation in input size.

\section{Conclusions and Future Work}
We demonstrated how Monte Carlo methods can be used to estimate runtimes of cryptographic and secure
multi-party computation protocols. We presented runtimes estimates, along with the respective
variances, for different dataset sizes. The simulation results were also used to fit regression
models with statistically significant coefficients showing the effect of input size on protocol
runtime.

We want to note that our implementations are very simple, with no optimizations of any kind. We are
only interested in validating the use of Monte Carlo integration for runtime estimation. Although the
observed values and estimations show a consistent advantage for the MPC protocol, it is important to
clarify that our experiments do not represent a very realistic scenario for MPC. As discussed in the
implementation details, the recorded runtimes do not include communication complexity and cost
introduced by distributing computation parties on multiple cloud providers or networks, since all
processes ran in threads on the same machine.

The impact of communication complexity on runtimes, as the number of computing parties grows, or
as other network variables (throughput, latency, etc.) vary, is a question open for further 
investigation.

\bibliographystyle{IEEEtran}
\nocite{*}
\bibliography{refs}

\end{document}